\documentclass[aps,floats,twocolumn,superscriptaddress]{revtex4}
\usepackage[final]{graphicx}
\DeclareGraphicsExtensions{.eps,.ps,.pdf}
\usepackage{amsmath,amssymb,bbold,bm}
\newcommand\vex[1]{\mathbf{#1}}

\def\d{\mathrm{d}}
\def\id{\mathbb{1}}

\def\ii{\mathrm{i}}
\def\RR{\mathrm{Re}}
\def\II{\mathrm{Im}}

\def\sslash#1{\setbox0=\hbox{$#1$}			
   \dimen0=\wd0                                		
   \setbox1=\hbox{/} \dimen1=\wd1  	 		
   \ifdim\dimen0>\dimen1                               	
      \rlap{\hbox to \dimen0{\hfil / \hfil}} 	  	
      #1                                      			
   \else                                        			
      \rlap{\hbox to \dimen1{\hfil$#1$\hfil}}   		
      \hbox{/} 	                              			
   \fi}   
\def\slash#1{\hbox{$#1$\kern-0.35em\raise0.1ex\hbox{/}}}

\begin{document}

\title{Vortices, zero modes and fractionalization in bilayer-graphene exciton condensate}
\author{B. Seradjeh}
\affiliation{Department of Physics and Astronomy, University of British Columbia,
Vancouver, BC, Canada V6T 1Z1}
\author{H. Weber}
\affiliation{Department of Physics and Astronomy, University of British Columbia,
Vancouver, BC, Canada V6T 1Z1}
\affiliation{Institut f\"ur Theoretische Physik, Universit\"at zu K\"oln, Z\"ulpicher Str. 77, 50937 K\"oln, Germany}
\author{M. Franz}
\affiliation{Department of Physics and Astronomy, University of British Columbia,
Vancouver, BC, Canada V6T 1Z1}
\date{\today}

\begin{abstract}
A real-space formulation is given for the recently discussed exciton condensate in a symmetrically biased graphene bilayer. We show that in the continuum limit an oddly-quantized vortex in this condensate binds exactly one zero mode per valley index of the bilayer. In the full lattice model the zero modes are split slightly due to intervalley mixing. We support these results by an exact numerical diagonalization of the lattice Hamiltonian. We also discuss the effect of the zero modes on the charge content of these vortices and deduce some of their interesting properties.
\end{abstract}

\maketitle

\emph{Introduction}.---Nontrivial topological configurations play a significant role in our understanding of a wide range of physical systems and often have interesting and potentially useful properties. In one spatial dimension, solitons and instantons interacting with fermions often induce fractional fermion numbers~\cite{JacReb76a,GolWil81a} as manifested in the canonical example of the polyacetylene chain \cite{SuSchHee79a}. In two dimensions, an important example is provided by vortices in a superfluid or a superconductor, precisely quantized to make possible the coexistence of the condensate with supercritical macroscopic rotation or magnetic field, respectively. Such vortices may carry fermonic zero modes which typically imply a host of unusual properties. In a planar spin-polarized $p+\ii p$ superconductor half-quantum vortices can exist and have been argued to obey non-Abelian exchange statistics due to the presence of an unpaired Majorana fermion bound to the vortex~\cite{Read1,Iva01a}. More recently similar physics has been shown to occur in certain simple models describing fermions on the honeycomb and $\pi$-flux square lattices where vortices in a bond dimerization pattern carry fractional charge~\cite{HouChaMud07a,SerWeeFra08a} and obey Abelian fractional statistics~\cite{Chamon2,SerFra07a}.

Topological structures in the above 2D examples and their associated phenomena afford a special degree of protection against local perturbations and have been identified as the key resource for fault-tolerant quantum information processing~\cite{Kit03a}. It would be clearly desirable if such systems could be produced and studied in the laboratory. Unfortunately, prospects for the experimental realization of the above systems are not particularly good: no spin polarized $p+\ii p$ superconductors are known to exist and it is not quite clear how one would realize the dimerization patterns required in Refs.~\cite{HouChaMud07a,SerWeeFra08a}.  

\begin{figure}[t]
\includegraphics[scale=0.6]{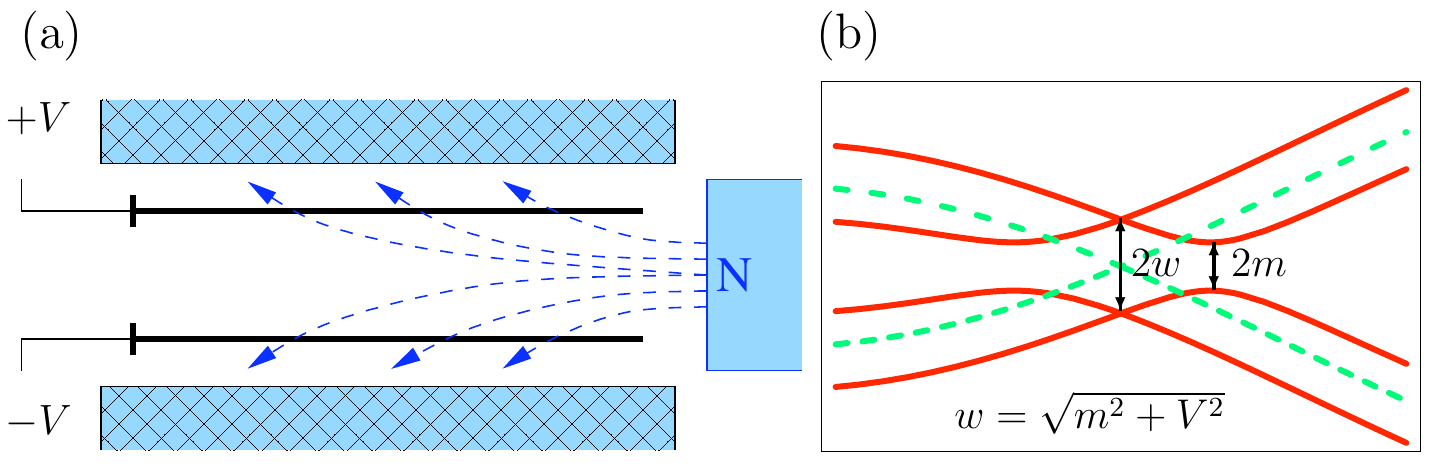}
\caption{(Color online)  (a) The schematic structure of the system, and the proposal to create and effective axial magnetic field. (b) The spectrum near the nodal point where the dashed (green) line shows the non-interacting, and  the solid (red) line the interacting case with the shifted nodes in external potential ($V$) and the exciton gap ($m$) marked.}\label{fig:1}
\end{figure}

In this Letter we discuss a physical system that is likely to be produced in a laboratory in the near future and show that it exhibits some of the key features of the aforementioned 2D models  ~\cite{Iva01a,HouChaMud07a,SerWeeFra08a,Chamon2,SerFra07a}.  Specifically, we consider a graphene bilayer separated by a dielectric barrier~\cite{MinBisSu08a} depicted in  Fig.~\ref{fig:1}(a). When biased by external gates the perfect nesting of the electron and hole Fermi surfaces in different layers creates ideal conditions for the formation of an exciton condensate (EC) i.e. the coherent liquid of bound states of electrons in one layer and holes in the other. It has been argued recently that for realistic values of parameters the EC can exist even at room temperature~\cite{MinBisSu08a}, and that it depends only weakly on the exact stacking of the two layers~\cite{ZhaJog08a}.  In what follows we construct a real-space model for this system and use it to study the internal structure of a vortex in the complex order parameter characterizing the EC. We show that it contains two fermion zero modes, one for each valley index of the bilayer, which are split slightly due to inter-valley mixing. Although there is no net charge associated with a vortex at half filling, the vortex binds fractional ``axial charge'', defined as the difference of the charge between the layers. We argue that such a vortex will obey fractional exchange statistics and comment on possible experimental signatures of our findings.

\emph{Exciton condensate}.---We consider the simplest model containing the essential physics of the system at zero temperature. It consists of fermions hopping between nearest neighbor (n.n.) sites on each of two layers of honeycomb lattice stacked directly on top of each other. Since the electron spin plays no role in the formation of EC we consider only a single spin projection. There is no direct hopping between the two layers. EC formation is driven by the interlayer Coulomb repulsion which we model by an effective short-range repulsion between the fermions in different layers at the same planar site. As argued prevoiusly ~\cite{MinBisSu08a} a layer of graphene is in the semi-metallic phase, meaning that the intra-layer Coulomb interaction is below the critical value needed for charge ordering. The long-range part of the interlayer Coulomb repulsion should only enhance the formation of the EC.

The Hamiltonian is, thus, $H=H_1+H_2+H_U$, where the in-plane Hamiltonian $H_\alpha=-t\sum_{\left<ij\right>}c^\dagger_{i\alpha} c^{}_{j\alpha}-(-)^\alpha V\sum_i n_{i\alpha}$ for layers $\alpha=1,2$ and the interaction $H_U=U\sum_i n_{i1}n_{i2}$. Here $c_{i\alpha}$ annihilates an electron at site $i$ and layer $\alpha$ and $n_{i\alpha}=c^\dagger_{i\alpha}c^{}_{i\alpha}$.

The mean-field order parameter of the EC is $\Delta_i=U\langle c^\dagger_{i2}c^{}_{i1}\rangle$. Decoupling $H_U$ in this channel we find 
\begin{equation}\label{eq:HMF}
H_{\mathrm{MF}}=H_1+H_2-\sum_i\left(\Delta_ic^\dagger_{i1}c^{}_{i2}+\mathrm{h.c.}\right)+\frac1U\sum_i|\Delta_i|^2.
\end{equation}
We now define $\Delta_\pm=\frac12(\Delta_{A}\pm\Delta_{B})$ where $A$ and $B$ are  the two sublattice sites in the same unit cell, and the spinor field operator $\psi_i=(c_{B_i1}, -c_{A_i1}, c_{B_i2}, c_{A_i2})^T$. For a uniform $\Delta_\pm$ we may write the Hamiltonian compactly in momentum space as $H=\sum_{\vex k}\psi^\dagger_{\vex k}h_{\vex k}\psi_{\vex k}^{}+E_0$, with $E_0=(N/U)(|\Delta_+|^2+|\Delta_-|^2)$, $N$ the number of sites per layer, and 
\begin{eqnarray}
h_{\vex k} = \gamma_0\bigg[\gamma_1  \RR\,t_{\vex k}+\gamma_2 \II\,t_{\vex k}+V\gamma_0\gamma_5+|\Delta_-|e^{-\ii\gamma_5\chi_-}\nonumber\\
 \ \ \ -\ii\gamma_1\gamma_2|\Delta_+|e^{-\ii\gamma_5\chi_+}\bigg].
\end{eqnarray}
Here $t_{\vex k}
=-t\sum_{\vex a\in\mathrm{n.n.}}e^{\ii\vex k\cdot\vex a}$;  
$\gamma_\mu=\ii\sigma_2\otimes\sigma_\mu$, $\gamma_0=\sigma_1\otimes\id$ are Dirac matrices in the Weyl representation, $\gamma_5=-\ii\gamma_0\gamma_1\gamma_2\gamma_3=\sigma_3\otimes\id$, and we have used a polar representation $\Delta_\pm=|\Delta_\pm|e^{i\chi_\pm}$.

At nonzero $\Delta_-$  a finite energy gap,
$E_g\leq |\Delta_-|,$
opens at half filling with the maximum occurring for fixed $\Delta_-$ when $|\Delta_A|=|\Delta_B|$. The lowest ground state energy occurs when the gap is maximum. Since a finite $|\Delta_+|$ does not open a gap on its own we also expect the ordered state with $\Delta_+=0$ to set in first. This can be checked explicitly by a calculation of the critical interaction for $|\Delta_+|$, which, for small $V\ll t$, yields $U_{c+}\sim t^2/V\gg t$.

For $\Delta_+=0$, the spectrum is $\pm E_{\vex k \alpha}$,
\begin{equation}
E_{\vex k \alpha}=\sqrt{\left(|t_{\vex k}|-(-)^\alpha V\right)^2+|\Delta_-|^2}.
\end{equation}
This is shown in Fig.~\ref{fig:1}b. The gap equation reads $2N=U\sum'_{\vex k\alpha}E_{\vex k\alpha}^{-1}$ where the sum is over the occupied states. The critical value of the interaction is given by $2N/U_{c-}=\sum'_{\vex k\alpha}||t_{\vex k}|-(-)^\alpha V|^{-1}$. Around the nodes the right-hand side diverges logarithmically for $\alpha=2$. So, $U_{c-}=0$ as expected, in contrast to $U_{c+}$. The dependence of $|\Delta_-|$ on $U$ and $V$ can be found in the nodal approximation to be 
\begin{equation}
|\Delta_-|\approx 2 \sqrt{V \Lambda} \exp\left( -\frac{\sqrt{3}\pi t^2}{UV} \right),
\end{equation}
with an ultraviolet cut-off $\Lambda\simeq t$ and in the limit $|\Delta_-|\ll V \ll \Lambda\ll t^2/U$. The state with $\Delta_+=0$ and $\Delta_-\neq 0$ is the EC phase considered in Refs.\ \cite{MinBisSu08a,ZhaJog08a}.

\emph{Low-energy theory.}---At long wave-lengths the system is dominated by the excitations around the two inequivalent valleys $\vex K_\pm=(\pm 4\pi/3\sqrt3a,0)$ of the graphene layers, where $t_{\vex K_\pm+\vex p}\approx\pm p_x+\ii p_y$. (We have set the Fermi velocity $3ta/2=1$.) The linearized mean-field Hamiltonian for these excitations is $h_++h_-$ with $h_+=\gamma_1\gamma_3h_-\gamma_3\gamma_1\equiv\mathcal{H}$,
\begin{equation}\label{ham}
\mathcal{H}=\gamma_0\left(\gamma_1\hat p_x+\gamma_2\hat p_y+V\gamma_0\gamma_5 + |m|e^{-\ii\gamma_5\chi} \right),
\end{equation}
where $\hat{\vex p}=-\ii\vex\nabla$ is the momentum operator and we have replaced $\Delta_-\equiv m=|m|e^{\ii\chi}$ for ease of use. The mass $m$ is now also taken to vary with position. Since the valleys are decoupled in the low-energy theory we shall first study a single valley. We discuss the effect of intervalley mixing later and return to the full Hamiltonian in the numerics.

The spectrum of $\mathcal{H}$ is symmetric around zero. This is seen by noting that
\begin{equation}\label{eq:symmH}
\gamma_2\mathcal{H}^*\gamma_2=\mathcal{H},
\end{equation}
and $\gamma_2^2=-1$.
Thus, for every eigenstate $\psi_E$ of energy $E$ the state $\Omega\psi_E\equiv\gamma_2\psi_E^*$ is an eigenstate of energy $-E$. This symmetry is furnished by the \emph{antiunitary} operator $\Omega=\Omega^\dag=\gamma_2K$ where $K$ is the complex-conjugation operator: $\{\Omega,\mathcal{H}\}=0$. When $V=0$ formally, there is another \emph{unitary} operator, $\gamma_0\gamma_3$, that anticommutes with $\mathcal{H}$ (Ref.~\onlinecite{JacPi07a}). For $V\neq0$, in contrast, $\Omega$ is the only anticummuting operator.

\emph{Zero modes.}---We will now consider a vortex in the EC order parameter, i.e. $m(r\to\infty,\theta)=m_0e^{\ii n\theta}$ where $n\in\mathbb{Z}$ is the vorticity, and show that, within the low-energy theory, there is exactly one zero mode per valley bound to the vortex for odd $n$. 

Before presenting the explicit solution, we note the following. For $V=0$ Eq.\ (\ref{ham}) formally coincides with the Hamiltonian studied in Refs.~[\onlinecite{HouChaMud07a,JacPi07a,SerWeeFra08a}], and hence there are $|n|$ exact zero modes for $n$-fold vortex~\cite{JacRos81a}. We now imagine slowly turning on $V$ and note that far from the vortex center the spectrum must remain gapped. Owing to Eq.~(\ref{eq:symmH}), which requires a symmetric spectrum, we could at most split an even number of zero modes. Therefore, at least one zero mode must survive for odd $n$. Also, since the spectral symmetry for $V\neq0$ is generated by an antiunitary operator, similar to the situation with half-quantum vortices in a $p+\ii p$ superconductor, we do not expect more than one zero mode to survive~\cite{GurRad07a}. This is confirmed by our explicit solution below.

To find zero-mode solutions $\mathcal{H}\psi_0=0$ we limit ourselves for simplicity to the quantum limit, $m(r\neq0,\theta)=m_0e^{\ii n\theta}$. The case of a more general radially symmetric vortex could be treated similarly. In the zero-energy subspace, $[\mathcal{H},\Omega]=0$. As a result we may take $\Omega\psi_0=\lambda\psi_0$ where the phase $\lambda=1/\lambda^*$ is an ``eigenvalue'' of antiunitary $\Omega$. We can always choose such a $\psi_0$: if $\Omega\psi_0\neq\lambda\psi_0$ we can pick $\psi'_0=(1+\lambda^*\Omega)\psi_0$ for which $\Omega\psi'_0=\lambda\psi'_0$. Since for an ``eigenstate'' $\Omega\phi_\lambda=\lambda\phi_\lambda$ we have  $\Omega(e^{-{\ii\beta}/2}\phi_{\lambda})=(e^{\ii\beta}\lambda)(e^{-{\ii\beta}/2}\phi_\lambda)$ we may further limit ourselves to $\lambda=1$. Such eigenstates are of the form $\psi_0=(f,g,\ii g^*,-\ii f^*)^T$ and form an over-complete basis of the spinor space.

With this preparation we find two independent equations for our zero modes,
\begin{subequations}
\begin{eqnarray}
e^{\ii\theta}(\partial_r+{\ii}r^{-1}\partial_\theta)f-m_0e^{\ii n\theta}f^*-\ii Vg &=& 0, \\
e^{-\ii\theta}(\partial_r-{\ii}r^{-1}\partial_\theta)g+m_0e^{\ii n\theta}g^*-\ii Vf &=& 0.
\end{eqnarray}
\end{subequations}
Any constant phase $e^{\ii\beta}$ of $m_0$ can be absorbed by $(f,g)\to e^{\ii\beta/2}(f,g)$ so we take $0<m_0\in\mathbb{R}$. We now make a ``one-phase'' \emph{ansatz} $f(r,\theta)=F(r)e^{\ii a\theta}$ and $g(r,\theta)=G(r)e^{\ii b\theta}$. We have checked that a ``two-phase'' \emph{ansatz} like the one used in Ref.~\onlinecite{JacRos81a} gives no new solutions when $V\neq0$. Thus, we are left with at most a single zero mode. We may eliminate the angular dependence by taking $a=b-1=\frac{n-1}2$. We see that when $n$ is even there is no single-valued solution. The radial part yields the bound-state Bessel-function solution
\begin{equation}
G(r) = e^{-m_0r}J_b(Vr),\quad F(r)=-\ii e^{-m_0r} J_a(Vr).
\end{equation}
For $V\to0$ and $n=\pm1$ we recover the zero mode of Ref.~\cite{HouChaMud07a} which has support only on one sublattice in each layer. Since we have at most a single zero mode, the $V\to0$ limit when $|n|>1$ is singular. 

A localized zero-mode in a gapped, particle-hole symmetric system is known to carry a fractional charge $\pm e/2$ \cite{JacReb76a,GolWil81a,SuSchHee79a,HouChaMud07a,SerWeeFra08a}. Since we have two independent valleys, however, we expect two zero modes to exist for odd $n$. Thus, charge fractionalizion should not occur. Nevertheless, we show below that the zero modes, which are in addition split slightly due to intervalley mixing, lead to fractional ``axial charge'' bound to a vortex. This axial charge is experimentally measurable and also endows the vortex with fractional exchange statistics. In the real system we must also include the spin of electrons. However, unlike the valley degree of freedom the vortex need not mix spins since in principle we can have a vortex in just one spin species.

\begin{figure}[t]
\includegraphics[scale=1.1]{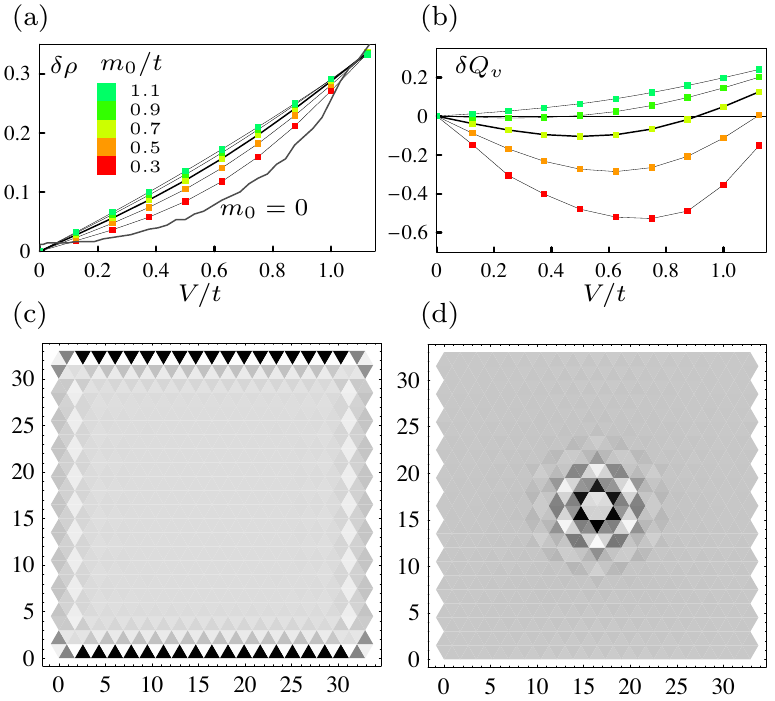}
\caption{(Color online) The ``axial charge''. (a) Density in the uniform system, and (b) the extra charge bound to a single vortex as a function of $V$ and $m_0$. The corresponding density profile for $V/t=0.4$ and $m_0/t=0.3$ in (c) the uniform system, and (d) the system with a vortex at the center. In all plots, each layer has $39\times22$ sites with a lattice size $\approx (35)^2$. The lattice sites in (c) and  (d) are at the center of the triangles.}\label{fig:2}
\end{figure}

\emph{Numerics.}---We have performed exact diagonalization of the mean-field Hamiltonian~(\ref{eq:HMF}) on honeycomb lattices with up to $51\times30$ sites per layer. (A layer with $N_x\times N_y$ sites has spatial dimensions $\frac{\sqrt3}2N_xa\times\frac32N_ya$.) 
Our numerical results support the conclusion that near-zero modes exist only for odd vorticity.

At half filling the charge is balanced between the layers, so that the total charge is always uniform, with or without a vortex. However, the charge difference, $\delta Q$, between the layers, which we call the ``axial charge'' in analogy with Ref.~\onlinecite{SerFra07a}, shows interesting features. (The axial charge is proportional to the electric dipole between the layers.) In the uniform system and at $m_0=0$ we have the axial density $\delta\rho\equiv\delta Q/N=\frac1N\sum_{|t_{\vex k}|<V}1\approx \frac1{\sqrt3\pi}(V/t)^2$ for $V\ll t$. Fig.~\ref{fig:2} shows our numerical results for the axial charge of a typical system with $39\times22\times2$ sites. In the uniform system, we observe a nearly linear dependence of $\delta\rho$ on $V<m_0$ which turns into quadratic for $V>m_0$. The density profile in Fig.~\ref{fig:2}(c) shows oscillations around the edges and dependence on the type of the edge (zig-zag for the horizontal and arm-chair for the vertical edges). 

A vortex sitting at the center of the system binds an \emph{irrational fraction} of the axial charge, $\delta Q_v$, on top of the uniform background. The dependence of $\delta Q_v$ on $V$ and $m_0$ in Fig.~\ref{fig:2}(b) shows that the value as well as the sign of this extra axial charge may be selected continuously by tuning external parameters. The density profile of this extra axial charge follows the shape of the
zero-mode wavefunction and is displayed in Fig.~\ref{fig:2}(d).

\begin{figure}[t]
\includegraphics[scale=0.95]{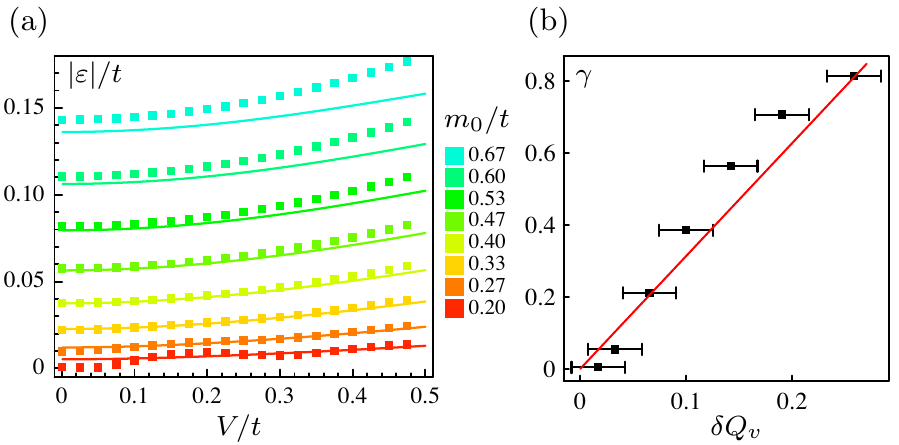}
\caption{(Color online) (a) The zero-mode energy splitting. The squares show our numerical results for a system size $\approx (35)^2$. The solid lines show the fitting with Eq.~(\ref{eq:split}). (b) The exchange phase vs. axial charge. The solid (red) line is  $\gamma=\pi\delta Q_v$. 
}\label{fig:3}
\end{figure}

Fig.~\ref{fig:3} shows the energy splitting of the zero modes obtained numerically. We can understand this analytically by evaluating the matrix element between the zero-modes obtained in the single-valley approximation and then diagonalizing the resulting $2\times 2$ Hamiltonian in the zero-mode subspace. This yields the zero mode splitting $\pm|\varepsilon|$ with
\begin{equation}\label{eq:split}
\varepsilon=\ii m_0\frac{\int e^{-\ii\vex Q\cdot\vex r}\RR\left[e^{-\ii\theta}\left(f^2+g^2\right)\right]\d\vex r}{\int\left(|f|^2+|g|^2\right)\d\vex r},
\end{equation}
where $\vex Q$ is the inter-valley momentum. Eq.~(\ref{eq:split}) shows the same qualitative behavior as the numerical results. In order to obtain a reasonable quantitative fit we must replace $\vex Q$ with an ``effective'' internodal wavevector, $\eta^{-1}\vex Q$, which represents the effects of the full spectrum. (Note that $\vex Q$ is completely extraneous to the nodal approximation.) A good fit is obtained with $\eta\approx3.0$ for low to intermediate values of $m_0$ and $V$.

\emph{Fractional statistics.}---The vortex carries fractional axial charge $\delta Q_v$ and, upon performing a singular gauge transformation employed in Ref.~\onlinecite{SerFra07a}, it can be seen to carry a half of an axial flux quantum, defined as the difference between the effective magnetic fluxes piercing the two layers. Employing the standard argument~\cite{wilczek_book} such charge-flux composite particles are expected to behave as Abelian anyons with the exchange phase $\gamma=\pi\delta Q_v$. In order to verify this hypothesis we have computed the exchange phase numerically using a generalized Bargmann invariant~\cite{simon}. Our results, summarized in Fig.~\ref{fig:3}(b), are consistent with the expected exchange phase, although system size limitations result in relatively large error bars.
It is important to emphasize that unlike the situation in the fractional quantum Hall fluids the value of the fractional charge and the exchange phase here is not protected by symmetry or topology but depends continuously on externally adjustable parameters. In this sense our situation is similar to ``irrational fractionalization'' of charge and statistics discussed in Ref.~\onlinecite{Chamon2}.

\emph{Discussion.}---The exciton condensate considered in this work should exhibit a host of unusual behaviors observable experimentally. The system is an insulator in the total charge channel but a superconductor in the axial channel. This behavior should lend itself to direct observation in transport if the two layers could be independently contacted. Such measurements are possible in bilayers fabricated using semiconductor structures~\cite{VigMac96a,KeoGupBee05a}.

The exciton order parameter has been shown to be minimally coupled to the ``axial gauge field'' i.e. the part of the vector potential ${\bf A}$ that produces the antisymmetric combination $\delta B = (\vex B_1-\vex B_2)_\perp$ of the magnetic field normal to the plane of the layers~\cite{BalJogLit04a}. Vortices in the EC therefore may be generated by the axial field just like Abrikosov vortices are generated by ordinary magnetic field in a superconductor. Indeed it is possible to arrange the external magnetic field in a way that $\delta B$ is much greater than the symmetric combination $(\vex B_1+\vex B_2)_\perp$. One such arrangement is schematically shown in Fig.~\ref{fig:1}(a). Using standard arguments unidirectional drift of vortices through the system can be seen to produce an axial Hall voltage from which the fractional axial charge carried by an individual vortex can be inferred. Such a drift can occur in response to the Magnus force acting on vortices in external applied axial current or, alternatively, due to a temperature gradient applied across the sample.

The zero modes found here exist in the excitonic instability of a Fermi surface. The existence of the underlying Fermi surface is at the root of excitonic instability at infinitesimal coupling and also causes the oscillatory Bessel-function behavior of our solution. Previous zero-mode solutions have, to the best of our knowledge, only appeared in systems with Dirac points. It is very interesting to see if zero modes of this type can be found in other physical systems. We also note that the U(1) vortices considered here are allowed by lattice symmetries, hence the issue of \emph{linear} confinement occuring in Refs.~\cite{HouChaMud07a,SerWeeFra08a} does not arise. Yet, vortices in the present model exhibit some of the same fascinating properties.

\emph{Acknowledgment}.---The authors have benefitted from useful correspondence with R. Jackiw. This research has been supported by NSERC, CIfAR
and the Killam Foundation. H.W. was supported by the DFG through the SFB 608 and the SFB/TR12, and the DAAD. 
 

\end{document}